\documentclass[12pt,reqno]{amsart}
\usepackage{amsmath,amssymb,amsfonts,amsthm}
\usepackage[mathscr]{eucal}
\usepackage[all]{xy}
\usepackage{hyperref}
\usepackage{setspace}
\usepackage{bm}
\usepackage{xcolor}
\usepackage{ytableau}
\allowdisplaybreaks

\author[Victoria Abakumova, Daniel Frolovsky, Hans-Christian Herbig, and Simon Lyakhovich]{Victoria Abakumova${}^\dagger$, Daniel Frolovsky${}^\dagger$, Hans-Christian Herbig${}^\ddagger$, and Simon~Lyakhovich${}^\dagger$}

\address{$\dagger$ Physics Faculty, Tomsk State University, Lenin ave. 36, Tomsk 634050, Russia \\ $\ddagger$ Universidade Federal do Rio de Janeiro,
Departamento de Matem\'atica Aplicada - Av. Athos da Silveira Ramos 149, Centro de Tecnologia - Bloco C - cep: 21941-909 - Rio de Janeiro - RJ - Brasil}

\email{abakumova@phys.tsu.ru, daniylfrolovsky@yandex.kz, herbighc@gmail.com, sll@phys.tsu.ru}

\title{Gauge symmetry of linearised Nordstr\"om gravity \\ and the dual spin two field theory}

\begin{document}

\maketitle
\begin{abstract}
  The field equations are proposed for the third rank tensor field with the hook Young diagram.
  The equations describe the irreducible spin two massless representation in any $d\geq 3$.
  The starting point of the construction is the linearised system of Einstein equations which includes the Nordstr\"om equation. This equation, being considered irrespectively to the rest of the Einstein system, corresponds to the topological field theory. The general solution is a pure gauge, modulo topological modes which we neglect in this article. We find the sequence of the reducible gauge transformations for the linearised Nordstr\"om equation, with the hook tensor being the initial gauge symmetry parameter. By substituting the general solution of the  Nordstr\"om equation into the rest of the Einstein's system, we arrive at the field equations for the hook tensor. The degree of freedom number count confirms, it is the spin two theory.
\end{abstract}

\section{Introduction}
The spin two massless irreducible representation admits, besides the linearised Einstein equations, alternative field-theoretical descriptions by the third rank tensor with hook Young diagram \cite{Curtright:1980yk}, \cite{Hull:2001iu}, \cite{Casini:2001gv}.  These alternative descriptions are connected with the symmetric tensor representation by the Hodge dualisations in $d\geq 5$, and they do not apply to $d\leq 4$. Being consistent at free level, these formulations are known to obstruct inclusion of interactions \cite{Bekaert:2002uh}. For various higher spin dualisations of this type, and review  of the recent developments, we refer to  \cite{Bekaert:2002dt}, \cite{Boulanger:2020yib}. We can also mention the recent work \cite{Krasnov:2021nsq} on the chiral formulation for higher spin fields. This formulation admits different higher spin interaction vertices comparing to the counterparts based on the symmetric tensors.

One more dual description of gravity worth
to mention is the Lanczos potential \cite{Lanczos:1949zz}, \cite{Lanczos:1962zz}, being also the third rank tensor with hook symmetry. The Lanczos tensor can be thought of as the potential  for the Weyl tensor \cite{Takeno}, by analogy with the vector potential for the electromagnetic strength tensor. This analogy arises
from the fact that the Bianchi identities, being imposed on the Weyl tensor $C$, imply that $C$ is a combination of the first derivatives of the third rank tensor with the hook symmetry. For discussion of physical interpretations of Lanzcos tensor and recent developments, we refer to \cite{Jezierski:2020ror}, \cite{Gopal:2021lax}. The Lanczos tensor is essentially $d=4$ structure \cite{Edgar:1997af}
 though the analogues with different Young diagrams exist in higher dimensions \cite{Edgar:2004iq}. One more somewhat similar dualisation scenario is proposed in reference \cite{Bergshoeff:2009tb} for the massive spin 2 field in $d=3$. In this work, the transversality equation for traceless second rank tensor field is solved by introducing the potential, which is also symmetric second rank tensor. In higher dimensions, the antisymmetric tensor potential is introduced for massive spin 1 in article \cite{Abakumova:2021evc} also by solving the transversality condition for the vector field.

In this article, we propose another alternative representation for the spin two by the third rank tensor with the hook Young diagram.
The key idea of this dualisation of the spin two field appeals to the same analogy with the equation $ dF = 0 $ for the spin one field strength, which motivates the introduction of the Lanczos potential for the Weyl tensor. We proceed from a different interpretation of the analogy, which leads to a different result.
The equation $dF=0$, being considered as such, irrespectively to the rest of Maxwell system, describes a topological field theory, in the sense that it has no degrees of freedom modulo De Rham cohomology.
The gauge symmetry  transformation for this equation is $\delta F = dA$, where $A$ is the gauge parameter, being an arbitrary one-form. The general solution is a pure gauge $F=dA$ if we neglect the topological degrees of freedom related to the De Rham cogomology. Substituting this solution into the remaining Maxwell equations $d*F=0$, one arrives at the equations for the potential $A$. 

Einstein's system without matter includes the Nordstr\"om equation
\begin{equation}\label{NE}
  R=-\,\frac{2d}{d-2}\Lambda\, ,
\end{equation}
where $R$ is the scalar curvature, $d$ is dimension of the space-time, and $\Lambda$ is the cosmological constant.
This equation, if considered by itself, independently of the entire Einstein system, appears to be a topological field theory, in the sense it does not have local degrees of freedom.
For the full non-linear theory (\ref{NE}), the complete set of infinitesimal gauge transformations is unknown in the explicit form, to the best of our knowledge. Diffeomorphism is the obvious gauge symmetry for the Nordstr\"om equation, though it is insufficient to gauge out all the local degrees of freedom of equation (\ref{NE}). Furthermore, the general solution for (\ref{NE}) cannot reduce to diffeomorphism, because this would mean Einstein's equations, having (\ref{NE}) among them, admit only trivial solutions. Implicitly, the existence of the extra gauge symmetry for eq. (\ref{NE}) is confirmed by the theorem due to Koiso \cite{Koiso}, see also \cite{Besse:1987pua}, which states that constant curvature metrics admit infinitesimal deformations. The existence theorem by Koiso does not provide the explicit form of the gauge symmetry transformation, however. For the linearised equation (\ref{NE}), we find the gauge symmetry in the explicit form in the next section. The gauge transformation $\delta h=\partial\cdot H$ is parameterised by the third rank tensor $H$ with the hook symmetry. This symmetry is reducible, and we find the complete sequence of the symmetry for symmetry gauge transformations. The local degree of freedom count (see in the Appendix) confirms that it is the topological field theory indeed. Hence, the general solution to the linearised equation (\ref{NE}) is the pure gauge $h=\partial\cdot H$, much like $F=dA$ is the general solution to the topological system $dF=0$ for the case of spin one. In this sense, the hook tensor serves as the potential for the metric. Substituting $h=\partial\cdot H$ into the linearised Einstein equations, we arrive at the equations for the hook $H$. Once the general solution of (\ref{NE}), being a subsystem of Einstein's equations, is substituted into the rest of the Einstein system, it should be an equivalent field theory. The degree of freedom count for the equations for the hook tensor confirms that we get the spin two theory. These equations for the hook tensor of spin two differ from the earlier known equations for dualisations of the metric, and also from the linearised equations for the Lanczos tensor. These equations are consistent in all $d\geq 3$, while the sequence of reducible gauge symmetry transformations depends on the dimension.

The article is organised as follows. In the next section, we find the reducible gauge symmetry of the linearised Nordstr\"om equation. In Section 3, we discuss the linearised Einstein equations reformulated in terms of the hook tensor. In Section 4, we discuss the results and further perspectives. The Appendix details the degree of freedom counting.

\section{Gauge symmetry of linearised Nordstr\"om equation}
For any system of Lagrangian equations, the Dirac-Bergmann algorithm allows one to find the gauge symmetry, in principle. For not necessarily Lagrangian equations, like  (\ref{NE}), the extension of the Dirac-Bergmann algorithm
is known \cite{Lyakhovich:2008hu}. There are two potential difficulties with the extended Dirac-Bergmann procedure of finding the gauge symmetry, given the non-Lagrangian field equation. First, the procedure implies to treat the equation as the evolutionary one. This assumes to split the space-time into the space and time, that breaks explicit general covariance, much like the Dirac-Bergmann method for variational equations. Second, the algorithm implies to invert the non-degenerate linear operators. In the context of field theory, these operators can be differential with respect to
the space coordinates. Inversion of these operators could result in the spacial non-locality of the infinitesimal gauge transformations though they would be local in time. For the $d=2$ field theories, the procedure of finding all the infinitesimal gauge symmetries in the local form is proposed in the article \cite{Lyakhovich:2013qfa}. For $d>2$, no systematic method is known to find the complete set of generators of local gauge symmetry, given a system of field equations. This explains why the complete gauge symmetry of the Nordstr\"om equation is still unknown.

For linear field equations, the problem of finding the gauge symmetry is much simpler than for non-linear systems.
For a systematic exposition of the covariant  procedure of finding the gauge symmetry of not necessarily Lagrangian linear systems, see Section 4 in the article \cite{Francia:2013sca}. Here, we apply this procedure to linearisation of equation (\ref{NE}).
Substituting into (\ref{NE}) decomposition of the metric $g_{\mu\nu}=\eta_{\mu\nu}+h_{\mu\nu}$  to the Minkowski background $\eta$ and the small deviation $h$, and taking the first order approximation in $h$, we arrive at the linearised Nordstr\"om equation
\begin{equation}\label{LNE}
  \hat{N}_{\mu\nu}(\partial)h^{\mu\nu}(x)=-\,\frac{2d}{d-2}\Lambda\,, \qquad \hat{N}_{\mu\nu}(\partial)=\partial_\mu\partial_\nu-\eta_{\mu\nu}\Box \,,
\end{equation}
where $\Box= \eta^{\mu\nu}\partial_\mu\partial_\nu$.
The most general infinitesimal gauge transformation of $h^{\mu\nu}$ reads
\begin{equation}\label{GT}
  \delta_{\mathcal{E}} h^{\mu\nu}=\hat{R}{}^{\mu\nu}{}_A(\partial)\mathcal{E}^A\, , \qquad A=1,\dots ,N,
\end{equation}
where the gauge generators $\hat{R}{}^{\mu\nu}{}_A(\partial)$ are supposed to be linear differential operators, while the gauge parameters $\mathcal{E}^A$ are assumed to be arbitrary real smooth functions of Minkowski space. The fact that $\hat{R}$ is the operator means that it is a homogeneous polynomial in $\partial_\lambda$ with constant real coefficients:
\begin{equation}\label{R-polynom}
  \hat{R}{}^{\mu\nu}{}_A(\partial)=R{}^{\mu\nu|\lambda_1\dots\lambda_k}{}_A\partial_{\lambda_1}\dots\partial_{\lambda_k}\,, \quad k=1,\ldots,n.
\end{equation}
The order $n$ of the differential operators $\hat{R}$, and their total number $N$ is unknown from an outset.
Once transformation (\ref{GT}) is supposed to be a symmetry of linearised Nordstr\"om theory, it has to leave
the l.h.s. of equation (\ref{LNE})  unchanged
\begin{equation}\label{Nord-gst}
  \hat{N}_{\mu\nu} (\partial) \delta_{\mathcal{E}} h^{\mu\nu}= 0\, , \quad\forall\, \mathcal{E}.
\end{equation}
As the gauge parameters $\mathcal{E}$ are arbitrary smooth functions, the above relation is equivalent to the relation for the gauge generators
\begin{equation}\label{EQR}
  \hat{N}_{\mu\nu} (\partial)\hat{R}{}^{\mu\nu}{}_A(\partial)=0\, .
\end{equation}
Given $\hat{N}_{\mu\nu}(\partial)$ (\ref{LNE}), this relation can be viewed as the equations defining the gauge symmetry generators $\hat{R}{}^{\mu\nu}{}_A(\partial)$.
The operators $\hat{R}_A$ are assumed to constitute the generating set for the solutions to the equations (\ref{EQR}) in the sense that any solution is spanned by these generators
  \begin{equation}\label{Complete}
  \hat{N}_{\mu\nu} (\partial) \hat{K}{}^{\mu\nu}(\partial) = 0\quad\Leftrightarrow\quad\exists\,\hat{K}{}^A(\partial):\, \hat{K}{}^{\mu\nu}(\partial)=\hat{K}{}^A (\partial)\hat{R}{}^{\mu\nu}{}_A(\partial)\, ,
\end{equation}
where the expansion coefficients $\hat{K}{}^A (\partial)$ are polynomials in $\partial_\lambda$.

Let us discuss the structure of eq. (\ref{EQR}).
The operator $\hat{N}_{\mu\nu}$ (\ref{LNE}) is an element of the $d(d+1)/2$-dimensional linear space of covariant symmetric tensors.
The gauge generators are elements of the dual space of contravariant symmetric tensors  orthogonal to $\hat{N}_{\mu\nu}$.
If the components of the tensor $\hat{N}_{\mu\nu}$ were the numbers, the orthogonal space would be $(d(d+1)/2-1)$-dimensional, and one could easily find the basis  of $d(d+1)/2-1$ independent vectors in this space. The subtlety is that the elements of $\hat{N}_{\mu\nu}$ are polynomials in $\partial_\lambda$, not numbers.  So, to find the gauge symmetry of the equation  (\ref{LNE}), one has to solve the equations (\ref{EQR}) with  respect to $\hat{R}$ being the elements of the ring of polynomials
$\mathcal{R}[\partial_0,\dots, \partial_{d-1}]$, where $\partial_\lambda$ are considered as commuting formal variables.
The problem of this type is known in commutative algebra as the issue of the first syzygy module \cite{E}. The generating set of the solutions to (\ref{EQR}) can be overcomplete, in the sense that the elements of any generating set $\hat{R}_A$ are linearly dependent with polynomial coefficients
\begin{equation}\label{R1--}
  \hat{R}{}^{\mu\nu}{}_A(\partial) \hat{R}{}^{(1)A}{}_{A_1}(\partial)=0\, .
\end{equation}
 From the algebraic perspective, this redundancy of the generating set of the first syzygy module corresponds to the second syzygy module \cite{E}.
 For the original gauge symmetry (\ref{GT}), the second syzygy means reducibility, i.e. the gauge parameters $\mathcal{E}^A$ admit the gauge transformations of their own such that do not affect the transformations of original fields $h^{\mu\nu}$:
 \begin{equation}\label{E1}
   \delta_{\mathcal{E}_1}\mathcal{E}^A=\hat{R}{}^{(1)A}{}_{A_1}\mathcal{E}^{A_1}\, , \qquad  \delta_{\mathcal{E}_1}\left( \delta_{\mathcal{E}}h^{\mu\nu}\right)=\hat{R}{}^{\mu\nu}{}_A\hat{R}{}^{(1)A}{}_{A_1}\mathcal{E}^{A_1}=0\,, \quad\forall\,\mathcal{E}^{A_1} \, .
 \end{equation}
  The second level gauge symmetry can be reducible again, that corresponds to the third syzygy module,  etc. In this way we arrive at the chain of syzygies
\begin{equation}\label{Rk--}
  \hat{R}{}^{(k-1)A_{k-1}}{}_{A_{k}}(\partial)\, \hat{R}{}^{(k)A_k}{}_{A_{k+1}}(\partial) =0\,, \quad k=1\,,\ldots\,, \quad A_0\equiv A.
\end{equation}
Once the chain of syzygies above is known, it defines the sequence of gauge-for-gauge symmetries such that the transformations of parameters of certain level do not contribute to the gauge transformation of previous level,
   \begin{equation}\label{Ek}
   \begin{array}{c}
   \delta_{\mathcal{E}_{k+1}}\mathcal{E}^{A_k}=\hat{R}{}^{(k+1)A_k}{}_{A_{k+1}}\,\mathcal{E}^{A_{k+1}}\, ,\\[3mm]
   \delta_{\mathcal{E}_{k+1}}\left(\delta_{\mathcal{E}_k}\mathcal{E}^{A_{k-1}}\right)= \, \hat{R}{}^{(k)A_{k-1}}{}_{A_{k}}\,\hat{R}{}^{(k+1)A_k}{}_{A_{k+1}} \mathcal{E}^{A_{k+1}}=0, \quad\forall\,\mathcal{E}^{A_{k+1}} \,,  \quad k=1,\ldots\,, \quad A_0\equiv A.
   \end{array}
    \end{equation}
  The choice of the generating set $\hat{R}$ is not unique for every syzygy module, and the length of the chain of syzygies can depend on the choice. The choice with the minimal length $L$ of the sequence of syzygies exists such that $L\leq d$, according to Hilbert Syzygy Theorem \cite{E}. Even though operator  $\hat{N}_{\mu\nu}$  of linearised Nordstr\"om equation (\ref{LNE}) is quadratic in $\partial_\lambda$, the generators of gauge symmetries could  be of any order. The examples of the higher order gauge symmetries in the field theory with the second order equations of motion can be found in \cite{Francia:2013sca}.  The complete sequence of the syzygies forms the resolution for the module. The minimal resolution (i.e. such that has the minimal length of the sequence of syzygies) is unique up to isomorphism.
As we see, the problem establishing the gauge symmetry for the linearised
Nordstr\"om equation  and the sequence of gauge-for-gauge symmetries, reduces to explicitly finding the minimal resolution for equations (\ref{EQR}). Technically this means, we have to find all the generators $\hat{R}$ by solving the sequence of equations (\ref{EQR}), (\ref{R1--}), (\ref{Rk--}), given the operator $\hat{N}$ (\ref{LNE}) involved in the first relation of the sequence. The solution is expected to be explicitly covariant, once the equation enjoys Poincar\'e symmetry.

Computer packages are known for finding the minimal resolutions.
To probe the possible structure of the resolution for eq. (\ref{EQR}), we have applied the Macaulay2 package \cite{M2}, which tells us that resolution is generated by the elements of the first order in $\partial_\mu$ in the dimensions $3, 4, 5$. The length of the resolution is $d-1$ in these cases. The computer output for the gauge generators is not explicitly covariant, so it cannot be immediately applied even in the lower dimensions, though it is helpful for restricting the covariant ansatz to solve the sequence of the equations (\ref{EQR}), (\ref{R1--}), (\ref{Rk--}).

Given the hint from computer calculations, we take the covariant first order ansatz for the gauge generators for the equation (\ref{LNE}), and the symmetries of all the higher levels.
 Once the symmetry is of the first order, the symmetry of the $k$-th level is parameterised by the tensor of rank $k+2$ as the original field is the second rank tensor.

Substituting the first order ansatz $\delta h^{\mu\nu}=\partial_\lambda R^{\mu\nu\lambda}$, where $R$ is an arbitrary tensor with symmetry\footnote{Symmetrization over a set of indices is denoted by round brackets, and anti-symmetrization by square brackets, e.g. $T^{(\mu\nu)\dots}=1/2(T^{\mu\nu \dots} + T^{\nu\mu \dots}),\,  T^{[\mu\nu]\dots}=1/2(T^{\mu\nu\dots} - T^{\nu\mu \dots})$. } in the first two labels, $R^{(\mu\nu)\lambda}=R^{\mu\nu\lambda}$,
into relations (\ref{Nord-gst}), we get the specific form of the equation (\ref{EQR}) that defines the gauge transformations (\ref{GT}). Solving equation
\begin{equation}\label{Nord-gst-R}
(\partial_\mu\partial_\nu-\eta_{\mu\nu}\square)\partial_\lambda R^{\mu\nu\lambda}=0\,,
\end{equation}
we arrive at the gauge transformation for the fields
 \begin{equation}\label{Hook1}
   \delta_Hh^{\mu\nu}=\partial_\lambda H^{\mu\nu\lambda}-\frac{1}{d-1}\eta_{\alpha\beta}(\eta^{\mu\nu}\partial_\lambda H^{\alpha\beta\lambda}+\partial^\nu H^{\alpha\beta\mu}+\partial^\mu H^{\alpha\beta\nu})\,,
 \end{equation}
with the gauge parameter $H^{\mu\nu\lambda}$ being arbitrary third rank tensor with the hook symmetry\footnote{We use the symmetric basis for the hooks.},
 \begin{equation}\label{Hook-sym}
H^{(\mu\nu)\lambda}=H^{\mu\nu\lambda}\, , \qquad H^{(\mu\nu\lambda)}=0\,.
\end{equation}
This tensor  is described by the Young diagram
\begin{equation*}
\begin{ytableau}
\mu & \nu  \\ \lambda  \
\end{ytableau}\,\,.
\end{equation*}
Using the first order ansatz $\delta H^{\mu\nu\lambda}=\partial_\rho R^{\mu\nu\lambda\rho}$  for the gauge symmetry of the gauge parameters with $R^{(\mu\nu)\lambda\rho}=R^{\mu\nu\lambda\rho}$, $R^{(\mu\nu\lambda)\rho}=0$, we arrive at the equation
\begin{equation}
\partial_\rho\partial_\lambda\big[R^{\mu\nu\lambda\rho}-\frac{1}{d-1}\eta_{\alpha\beta}\big(\eta^{\mu\nu}R^{\alpha\beta\lambda\rho}+\eta^{\nu\lambda}R^{\alpha\beta\mu\rho}+\eta^{\lambda\mu}R^{\alpha\beta\nu\rho}\big)\big]=0\,.
\end{equation}
The solution to this equation defines gauge symmetry transformations (\ref{E1}) for the hook. This transformation reads
\begin{equation}\label{Hook2}
   \displaystyle \delta_{H_1}H^{\mu\nu\lambda}=\partial_\rho\big(H^{\mu\nu\lambda\rho}-\frac{1}{3}\frac{1}{d-1}\eta_{\alpha\beta}(2\eta^{\mu\nu}H^{\alpha\beta\lambda\rho}-\eta^{\nu\lambda}H^{\alpha\beta\mu\rho}-\eta^{\lambda\mu}H^{\alpha\beta\nu\rho})\big)\,,
\end{equation}
where $H^{\mu\nu\lambda\rho}$ is a tensor with hook symmetry type,
\begin{equation*}
\begin{ytableau}
\mu & \nu  \\ \lambda \\ \rho \
\end{ytableau}\,\,,
\end{equation*}
i. e. $H^{(\mu\nu)\lambda\rho}=H^{\mu\nu\lambda\rho}$, $H^{(\mu\nu\lambda)\rho}=0$, $H^{\mu\nu(\lambda\rho)}=0$.
Along the same line, we arrive to the following answer for sequence of gauge transformations (\ref{Ek}):
\begin{equation}\label{Hookk-}
   \delta_{H_k}H^{\mu\nu|\lambda\rho_1\dots\rho_{k-1}}=\partial_{\rho_k}H^{\mu\nu\lambda\rho_1\ldots\rho_k}\,, \qquad
   k=2,\ldots,d-2\,,
\end{equation}
with the gauge parameters being the tensors of rank $k+3$ at $k$-th level with the hook symmetry
\begin{equation*}
 \begin{ytableau}
\mu & \nu \\ \lambda \\ \rho_1 \\ \vdots \\ \rho_k \
\end{ytableau}\,\,.
\end{equation*}
Relations (\ref{Hook1}), (\ref{Hook2})--(\ref{Hookk-}) provide the complete sequence of the reducible gauge symmetry transformations for the original metric obeying the linearised Nordstr\"om equation, and for the gauge parameters of all the levels.

Once the gauge parameters are the hook-type tensors, the question can be asked about the linearised diffeomeorphism transformations, being the obvious gauge symmetry of the linearised Nordstr\"om
equation, as these transformations are parameterised by the vectors, not the hook-type tensors.
In fact, the diffeomorphism is included in the  transformation (\ref{Hook1}) parameterised by the traceful hook.
To see that, let us decompose the hooks into the trace and traceless part,
\begin{equation}\label{Htr}
\begin{array}{c}
\displaystyle H^{\mu\nu\lambda_1\ldots\lambda_s}=\widetilde{H}^{\mu\nu\lambda_1\ldots\lambda_s}+\frac{1}{2}\frac{1}{d-s}\Big(2\eta^{\mu\nu}H'^{\lambda_1\ldots\lambda_s}-s\big(\eta^{\nu[\lambda_1}H'^{\mu|\lambda_2\ldots\lambda_s]}+\eta^{[\lambda_1|\mu}H'^{\nu|\lambda_2\ldots\lambda_s]}\big)\Big)\,,
\\[3mm]
\eta_{\mu\nu}\widetilde{H}^{\mu\nu\lambda_1\ldots\lambda_s}\equiv 0\,, \qquad \eta_{\mu\nu}H^{\mu\nu\lambda_1\ldots\lambda_s}\equiv H'^{\lambda_1\ldots\lambda_s}\,, \qquad s=1,\ldots,d-1\,.
\end{array}
\end{equation}
Introducing separate notation for the traces with appropriate normalisation
\begin{equation}\label{vpar}
\displaystyle v^\lambda\equiv-\frac{3}{2}\frac{1}{d-1}H'^\lambda\,, \qquad  v^{\lambda\rho_1\ldots\rho_r}\equiv(-1)^{r+1}\frac{1}{2}\frac{1}{d-r-1}H'^{\lambda\rho_1\ldots\rho_r}\,, \qquad r=1,\ldots,d-2\,,
\end{equation}
we arrive to the gauge transformation that explicitly includes diffeomorphism in the familiar form,
\begin{equation}\label{Hook1tr}
\displaystyle \delta h^{\mu\nu}=\partial_\lambda\widetilde{H}^{\mu\nu\lambda}+\partial^\nu v^\mu+\partial^\mu v^\nu\,.
\end{equation}
Decomposing the hook tensors into the trace and traceless part, we can reorganise the sequence of gauge-for-gauge transformations in the following way
\begin{eqnarray}\notag
\displaystyle \delta \widetilde{H}^{\mu\nu\lambda_1\ldots\lambda_r}&=&
\partial_{\lambda_{r+1}}\Big(\widetilde{H}^{\mu\nu\lambda_1\ldots\lambda_{r+1}}+
(-1)^{r+1}\frac{1}{d-r}\big(2\eta^{\mu\nu}v^{\lambda_1\ldots\lambda_{r+1}}-
r(\eta^{\nu[\lambda_1}v^{\mu|\lambda_2\ldots\lambda_r]\lambda_{r+1}}
\\[3mm]\label{Hookrtr}
\displaystyle &+&\,\eta^{[\lambda_1|\mu}v^{\nu|\lambda_2\ldots\lambda_r]\lambda_{r+1}})\big)\Big)+\partial^\nu v^{\mu\lambda_1\ldots\lambda_r}+\partial^\mu v^{\nu\lambda_1\ldots\lambda_r}\,,\\[3mm]\label{vr}
\displaystyle \delta v^{\lambda_1\ldots\lambda_r}&=&-\frac{d-r-1}{d-r}\partial_{\lambda_{r+1}}v^{\lambda_1\ldots\lambda_{r+1}}\,, \qquad r=1,\ldots,d-2\,. \label{vrtr}
\end{eqnarray}
It is curious to notice that the transformations for the hook traces $v$, being antisymmetric tensors, decouple  from the transformations of the traceless hook parameters, while the transformations of $\widetilde{H}$ involve the higher order $v$'s. Let us elaborate on the structure of these gauge transformations.
At the level of original gauge transformation (\ref{Hook1tr}) both the traceless hook parameter and the vector are involved in a homogeneous way, with $v^\mu$ inducing the diffeomorphism. This vector gauge parameter enjoys the gauge symmetry parameterised by bi-vector $v^{\mu\nu}=-\,v^{\nu\mu}$, $\displaystyle\delta v^\mu =-\frac{d-2}{d-1}\partial_\nu v^{\mu\nu}$. The shift is transverse, $\partial_\mu \delta v^\mu=0$, so the volume preserving diffeomorphism may seem gauged out from the entire deffeomorphism algebra. In fact, the situation is different: the volume preserving diffeomorphism subalgebra can be absorbed by the transformations generated by traceless hook. In particular, taking the traceless hook of the special form
\begin{equation}\label{tildeH-v}
\widetilde{H}^{\mu\nu\lambda}=\partial_\rho\Big(\widetilde{H}^{\mu\nu\lambda\rho}+\frac{1}{d-1}\big(2\eta^{\mu\nu}v^{\lambda\rho}-\eta^{\nu\lambda}v^{\mu\rho}-\eta^{\lambda\nu}v^{\nu\rho}\big)\Big)+\partial^\nu v^{\mu\lambda}+\partial^\mu v^{\nu\lambda}\,,
\end{equation}
and substituting that into the gauge transformation (\ref{Hook1tr}), we get the diffeomorphism  $\delta h_{\mu\nu}\sim\partial_\mu v_\nu+\partial_\nu v_\mu$ induced by transverse vector $v^\mu\sim\partial_\nu v^{\mu\nu}$. In this sense, the volume preserving diffeomorphism is included into the gauge transformation (\ref{Hook1tr}) twice: for the first time as a part of the usual diffeomorphism transformation, and for the second time as the transformation (\ref{Hook1tr}) induced by the hook of the special form (\ref{tildeH-v}). These two sources of the volume preserving diffeomorphism can compensate each other, so the original field can remain unchanged under the transformation (\ref{Hook1tr}). It is the source of reducibility of gauge symmetry related to the volume preserving diffeomorphism. Once the volume preserving diffeomorphisms are the reducible gauge transformations by themselves, this leads to the full sequence of gauge-for-gauge transformations (\ref{vrtr}). From the perspective of Riemannian geometry, the source of the gauge symmetry (\ref{Hookrtr}) parameterised by the hook tensors is not evident at the moment.

To finalise this Section, let us mention that one can count degree of freedom number in explicitly covariant way by the recipe of the article \cite{Kaparulin:2012px}, given the field equations and their gauge symmetry, including the symmetry for symmetry. For the linearised Nordstr\"om equation this count is detailed in the Appendix. The result is that the Norsdtr\"om equation does not have local degrees of freedom indeed.

\section{The field equations for the hook tensor representing massless spin two}
In this section, we use the complete gauge symmetry of the linearised Nordstr\"om equation to construct the dual formulation of the massless spin 2 theory.

We begin with the Lagrangian of linearised Einstein gravity with cosmological constant $\Lambda$,
\begin{equation}\label{LagrLG}
\displaystyle \mathcal{L}=\frac{1}{4}\big(\partial_\mu h_{\nu\lambda}\partial^\mu h^{\nu\lambda}+2\partial^\mu h\partial^\nu h_{\nu\mu}-2\partial^\mu h^{\nu\lambda}\partial_{\lambda}h_{\nu\mu}-\partial_\mu h\partial^\mu h\big)-\Lambda h\,, \qquad h\equiv\eta^{\mu\nu}h_{\mu\nu}\,.
\end{equation}
The corresponding Lagrange equations read
\begin{equation}\label{LELGh}
\displaystyle L_{\mu\nu}\equiv\frac{1}{2}\big(\partial_\mu\partial^\lambda h_{\nu\lambda}+\partial_\nu\partial^\lambda h_{\mu\lambda}-\square h_{\mu\nu}-\partial_\mu\partial_\nu h\big)-\frac{1}{2}\eta_{\mu\nu}\big(\partial^\lambda\partial^\rho h_{\lambda\rho}-\square h\big)-\Lambda\eta_{\mu\nu}=0\,.
\end{equation}
The Nordstr\"om equation is the trace of the Einstein system. If the Nordstr\"om equation is considered separately from the entire Einstein system, it will be the field without local degrees of freedom. Hence, the general solution to the Nordstr\"om equation is a gauge transformation of any particular solution. Given the gauge symmetry transformations (\ref{Hook1}), the general solution reads
\begin{equation}\label{hHx}
\begin{array}{c}
\displaystyle h^{\mu\nu}=\partial_\lambda H^{\mu\nu\lambda}-\frac{1}{d-1}\big(\eta^{\mu\nu}\partial_\lambda H'^\lambda+\partial^\nu H'^\mu+\partial^\mu H'^\nu\big)\\[3mm]
\displaystyle+\,\frac{2\Lambda}{(d-2)(d-1)(2k-1)}(x_\mu x_\nu+k\eta_{\mu\nu}x_\lambda x^\lambda)\,,
\end{array}
\end{equation}
where proportional to $\Lambda$ term is the particular solution, and  $k=const$. This particular solution can be viewed as the small $\Lambda$ approximation to the constant curvature space metric in Lorentzian coordinates. We do not elaborate on this interpretation here, the details can be found in the article \cite{AL2021}.
Substituting the metric (\ref{hHx}), being the general solution of linearised Nordstr\"om equation into linearised Einstein equations (\ref{LELGh}), we arrive at the equations for the hook tensor,
\begin{equation}\label{LELGH}
\begin{array}{c}
\displaystyle E_{\mu\nu}(H)\equiv\frac{1}{2}\big(\partial_\mu\partial^\lambda\partial^\rho H_{\nu\lambda\rho}+\partial_\nu\partial^\lambda\partial^\rho H_{\mu\lambda\rho}-\square\partial^\lambda H_{\mu\nu\lambda}\big)\\[3mm]
\displaystyle-\,\frac{1}{2}\frac{1}{d-1}
\big(\partial_\mu\partial_\nu\partial_\lambda H'^\lambda-\eta_{\mu\nu}\square\partial_\lambda H'^\lambda\big)
=0.
\end{array}
\end{equation}
Once the general solution of the subsystem is substituted back in the entire system, this does not change dynamical content of the theory.
Because of this reason, the equations for the hook (\ref{LELGH}) have to be equivalent to the original linearised Einstein equations. The hook can be viewed, in a sense, as a potential for the metric of constant scalar curvature, as it is discussed in the Introduction.

Let us now discuss the gauge algebra of the equations for hook.
By construction, equations (\ref{LELGH}) have  the same gauge identities among them as the Einstein system,
\begin{equation}\label{gidLGH}
\displaystyle \displaystyle \partial^\nu E_{\mu\nu}(H)\equiv0\,.
\end{equation}
The sequence of gauge symmetries for field equations (\ref{LELGH}) coincides with (\ref{Hook2})--(\ref{Hookk-}), as these transformations do not affect the metric (\ref{hHx}).

The system (\ref{LELGH}) is obviously non-Lagrangian as the l.h.s. is the symmetric second rank tensor, while the field is the third rank tensor with the hook Young diagram. From this perspective, it seems natural that no pairing is seen between the gauge symmetries and gauge identities as the second Noether theorem does not apply to the non-Lagrangian systems. Be the system Lagrangian or not, one can count the degree of freedom (DoF) number, given the gauge symmetries and gauge identities of the field equations. This count is detailed in the Appendix.
The DoF number is $d^2-3d$ by phase-space count that confirms once again that equations (\ref{LELGH}) indeed describe the massless spin two.

In the end of this section let us mention that equations (\ref{LELGH}) for the spin two in the hook representation do not explicitly involve the cosmological constant, while linearised Einstein's system (\ref{LELGh}) includes $\Lambda$-term. This does not indicate inequivalence because at the linearised level the constant source can be always excluded by the linear local redefinition of the fields, even directly in the Lagrangian. To get rid of $\Lambda$ in the linear equations, one just shifts the metric $h_{\mu\nu}$ by the particular solution of the equation with specific constant in the right hand side. Beyond the linear level, the cosmological constant has to be accounted for indeed.

\section{Concluding remarks}
 In conclusion, let us discuss the two main results of the article.

The first result is that complete reducible gauge symmetry of the Nordstr\"om equation is found at linearised level. Given the gauge symmetry, we proof that the theory is topological in the sense that all the local degrees of freedom are gauged out. Beyond the linearised level, the complete gauge symmetry of Nordstr\"om equation still remains an open issue. The non-linear theory is obviously consistent, as the solutions exist to the field equation. The Nordstr\"om equation admits smooth linearisation, so one can think that the gauge symmetry of linear approximation can be consistently deformed in
 a perturbative way. One more argument supporting the existence of gauge symmetry of the non-linear Nordstr\"om equation, besides the diffeomorphism,  is the Koiso theorem \cite{Koiso}, \cite{Besse:1987pua}. The theorem states that constant scalar curvature spaces admit infinitesimal deformations in the vicinity of general metric with $R=\Lambda$. It is the existence theorem that does not provide explicit form for the gauge generators. The theorem also states that exceptional constant scalar curvature metrics are possible such that existence of the deformation is in question. These exceptions occur when $\Lambda$ is correlated in certain way with eigenvalues of Laplacian for the space with this metric. This can indicate that gauge symmetry of non-linear Nordstr\"om equation admits some stationary points in the space of metrics.  A toy model example for a similar phenomenon can be found in the article \cite{Lyakhovich:2014soa}. It is the example of a simple topological field theory where the gauge symmetry admits stationary points in the space of fields.  As one can learn from this example, the linear limit of the gauge symmetry is reducible, and it still has stationary points, while the linear limit of the field equation (the limit is smooth) admits irreducible gauge symmetry without stationary points. This can indicate that problem of finding the gauge symmetry for Nordstr\"om equation beyond the linearised level is not necessarily solvable by deforming the transformations of linear theory.

The second result of the article is the dual representation for the massless spin two theory by the third rank tensor with the hook Young diagram subject to equations (\ref{LELGH}). This result is deduced from the previous one. Once the linearised Nordstr\"om equation is a topological field theory, the general solution is a pure gauge. In this sense the hook is a potential for Nordstr\"om metric. Given the metric in terms of the hook potential, we substitute it into the  linearised Einstein system arriving at equations (\ref{LELGH}). By construction, these equations are equivalent to the original equations for the symmetric second rank tensor.  Equations (\ref{LELGH}) enjoy reducible gauge symmetry which includes the volume preserving diffeomorphism transformations and some other symmetries whose geometric interpretation is not evident at the moment. Given the complete gauge symmetry, degree of freedom count confirms, it is the massless spin two theory.

Equations (\ref{LELGH}) for spin two are obviously non-Lagrangian. Making use of the general Stueckelberg scheme of the article \cite{Lyakhovich:2021lzy}, these equations can be cast into Lagrangian setup.
The method of including the Stueckelberg fields of the work \cite{Lyakhovich:2021lzy} begins with completion of the Lagrangian equations by the differential consequences. The Stueckelberg field is introduced for every added consequence. The outcome is proven equivalent to the original theory, and it is Lagrangian. If the set of added consequences is overcomplete, the  Stueckelberg symmetry is reducible \cite{Abakumova:2021evc}. For the problem at hands, Einstein's system is to be complemented by the third order consequence with the hook Young diagram. Then the method of article \cite{Abakumova:2021evc} is applied, and the result will be the equivalent Lagrangian theory involving simultaneously both metric and hook tensor.
The Stueckelberg gauge symmetry can be fixed in different ways. One way is to fix the hook that leads to the original Einstein system. Another gauge condition fixes the symmetric tensor reducing the system to equations (\ref{LELGH}). To
put it differently, the Lagrangian formulation is possible which simultaneously involves both metric and its potential, being the hook tensor. Imposing different gauges one can switch between these two dual formulations. Construction of this Lagrangian is work in progress. This construction seems working in unobstructed way at the non-linear level. The matter is that the non-linear Einstein's equations also admit the differential consequence with the hook Young diagram, and the Stueckelberg field can be iteratively included proceeding from this consequence in explicitly covariant way. If the original system is linearised, the inclusion of Stueckelberg fields terminates at squares.
In the non-linear
case it continues to the higher orders. It is worth to notice that the procedure of iterative inclusion of Stueckelberg fields is unobstructed if the original system is consistent \cite{Lyakhovich:2021lzy}. This can open the way to construct the dual theory of massless spin two which admits consistent interactions. Once the metric is switched off from the Stueckelberg formulation by gauge fixing (which reduces the metric to the fixed one), the hook self-interactions will explicitly depend on the background metric. This dependence does not seem essential at least for infinitesimal deformations of the background metric, as it is just another gauge in the Stueckelberg theory.

As the last remark let us mention that the higher derivatives in the field equations for hooks (\ref{LELGH}) do not mean instability.
Various higher derivative theories are stable, see e.g. \cite{Kaparulin:2014vpa}, \cite{Abakumova:2018eck}, \cite{Abakumova:2017uto}, even though the canonical energy is unbounded. This happens because there can exist the other bounded conserved quantity which stabilizes the dynamics. For the third order equations (\ref{LELGH}), the bounded conserved quantity is obvious: it is the canonical energy of the Einstein's linearised theory expressed in terms of the hook tensor.

\vspace{0.2 cm}
\subsection*{Acknowledgements} We thank A.A.~Sharapov for fruitful discussions. The part of the work concerning the study of gauge symmetry of Nordstr\"om equation is supported by Foundation for Advancement of Theoretical Physics and Mathematics ``Basis". The dual formulation of spin two theory is a part of the project supported by a government task of the Ministry of Science and Higher Education of the Russian Federation, Project No. FSWM-2020-0033.

\section*{Appendix \\Degree of freedom count}
\noindent
In this appendix, we at first explain the recipe for the degree of freedom (DoF) counting proposed in the article \cite{Kaparulin:2012px}.
For the sake of simplicity, the explanations are adjusted for the context of the problem addressed in the present work. In particular, we consider here only linear systems of field equations, while the article \cite{Kaparulin:2012px} deals with any field theory.
Then, making use of this recipe, we count the DoF for the Nordstr\"om equation and for the dual equations (\ref{LELGH}) for the spin two.

Consider the set of fields $\phi^i$ subject to a system of linear field equations,
\begin{equation}\label{LFE}
  T_{ai}(\partial)\,\phi^i=0 \,  .
\end{equation}
The matrix of the wave operator $T_{ai}(\partial)$ has the elements being polynomials in $\partial$.
Once the classical dynamics of the fields are completely defined by the field equations, the structure of polynomials $T_{ai}(\partial)$ encodes all the relevant data about the theory, including the DoF number. Let us explain how the DoF number can be extracted from the structure of the polynomials $T(\partial)$.

At first, the system (\ref{LFE}) is assumed \emph{involutive}, i.e. it should not admit the lower order differential consequences. Of course, not every reasonable system of linear field equations is involutive from the outset. For example, the second order Proca equations for the massive spin one field are not involutive as the first order consequence exists being the transversality condition,
\begin{equation}\label{Proca}
  \big( ( \Box - m^2 )\eta_{\mu\nu}-\partial_\mu\partial_\nu \big)A^\nu=0 \quad \Rightarrow\quad m^2\partial_\mu A^\mu=0 \, .
\end{equation}
If the original system is not involutive, it can be brought to the involution by inclusion of all the lower order consequences.
So, the assumption that the system (\ref{LFE}) is involutive does not restrict generality once any system can be equivalently reformulated in the involutive form. The DoF counting, however, proceeds from the involutive form of the equations.

Every equation (\ref{LFE}) labelled by the index $a$ has the order $t_a$  which is understood as the maximal order of the polynomials $T_{ai}$ for all $i$,
\begin{equation}\label{ordt}
\displaystyle t_a=\max_{i}\big\{\text{ord}(T_{ai})\big\}\,.
\end{equation}
For example, the involutive closure of Proca system  (\ref{Proca}) includes four equations of the second order (the Proca equations themselves) and one equation of the first order (the transversality condition).

The wave operator matrix $T_{ai}(\partial)$ can admit null-vectors from the left and from the right,
\begin{equation}\label{L1}
  L_A{}^a (\partial)\, T_{ai}(\partial)=0 \, ,
\end{equation}
\begin{equation}\label{R1}
  T_{ai}(\partial)  R^i{}_\alpha (\partial)\,=0 \, ,
\end{equation}
where $L, R$ are polynomials in $\partial$. We assume that $L_A{}^a$, $R^i{}_\alpha$ form the generating sets for the left and right kernel of the matrix  $T_{ai}(\partial)$ in the sense that any left- or right- null-vector is spanned by $L_A$ or $R_\alpha$ respectively, with the coefficients being polynomials in $\partial_\mu$.
Relations (\ref{L1}) mean the gauge identities between the field equations (\ref{LFE}), so $L$'s are considered as the generators of identities. Relations (\ref{R1}) mean the gauge symmetry transformations generated by $R$,
\begin{equation}\label{GS1}
\delta_{\mathcal{E}}\phi^i=R^i{}_\alpha(\partial)\,\mathcal{E}^\alpha, \qquad T_{ai}(\partial)\, \delta_{\mathcal{E}}\phi^i \equiv 0, \quad \forall\,\mathcal{E}^\alpha \, .
\end{equation}
Let us now explain how the order is computed of Noether identities and gauge symmetries. The order of the $a$-th identity component of (\ref{L1}) is defined as the sum of $t_a$ (\ref{ordt}) and the order of polynomial $L_A{}^a(\partial)$. Then the total order of the gauge identity generated by $L_A$ is defined as follows:
\begin{equation}\label{ordl1}
\displaystyle l_A=\max_a\big\{t_a+\text{ord}(L_A{}^a)\big\}\,.
\end{equation}
The total order of gauge symmetries (\ref{GS1}) is defined as the maximal order of the polynomials $R^{}_\alpha$ for all $i$,
\begin{equation}\label{ordr1}
\displaystyle r_\alpha=\max_{i}\big\{\text{ord}(R^i{}_\alpha)\big\}\,.
\end{equation}
For the Proca case, there exists a single gauge identity of the third order,
\begin{equation}
\displaystyle \partial^\mu\big[\big((\square-m^2)\eta_{\mu\nu}-\partial_\mu\partial_\nu\big)A^\nu\big]+m^2\partial_\mu A^\mu=0\,,
\end{equation}
and the system has no gauge symmetry.

The gauge identities (\ref{L1}) and gauge symmetry tranformations (\ref{R1}) can be, in general, reducible, i.e. there exist relations
\begin{equation}\label{L2}
\displaystyle L_{A_1}{}^A(\partial)L_A{}^a(\partial)=0\,,
\end{equation}
\begin{equation}\label{R2}
\displaystyle R^i{}_\alpha(\partial) R^\alpha{}_{\alpha_1}(\partial)=0\,,
\end{equation}
which can be further reducible,
\begin{equation}\label{Lk}
\displaystyle L_{A_{k_L}}{}^{A_{k_L-1}}(\partial)
L_{A_{k_L-1}}{}^{A_{k_L-2}}(\partial)=0\,,
\end{equation}
\begin{equation}\label{Rk}
\displaystyle R^{\alpha_{k_R-2}}{}_{\alpha_{k_R-1}}(\partial)R^{\alpha_{k_R}-1}{}_{\alpha_{k_R}}(\partial)=0\,,
\end{equation}
where $k_L=2,\ldots,m_L$, $A_0\equiv A$, and $k_R=2,\ldots, m_R$, $\alpha_0\equiv\alpha$.
The total orders of gauge identities and gauge symmetries of the $k$-th stage of reducibility are defined as follows:
\begin{equation}\label{ordlk}
\displaystyle l_{A_k}=\max_{A_{k-1}}\big\{l_{A_{k-1}}+\text{ord}(L_{A_k}{}^{A_{k-1}})\big\}\,, \quad k=1,\ldots\,, \quad A_0\equiv A\,,
\end{equation}
\begin{equation}\label{ordrk}
\displaystyle r_{\alpha_k}=\max_{\alpha_{k-1}}\big\{r_{\alpha_{k-1}}+\text{ord}(R^{\alpha_{k-1}}{}_{\alpha_k})\big\}\,, \quad k=1,\ldots\,,
\quad \alpha_0\equiv \alpha\,.
\end{equation}
The number of DoF for the field equations with reducible gauge symmetries and/or identities, can be found by the recipe
\begin{equation}\label{NDoF1}
\displaystyle \mathcal{N}=\sum\limits_{a}t_a-\sum\limits_{k}(-1)^k\Big(\sum\limits_{A_k}l_{A_k}+\sum\limits_{\alpha_k}r_{\alpha_k}\Big)\,.
\end{equation}
Here, $l_{A_k}$ are defined by (\ref{ordl1}) and (\ref{ordlk}), $A_0\equiv A$, and $r_{\alpha_k}$ are defined by (\ref{ordr1}) and (\ref{ordrk}), $\alpha_0\equiv \alpha$.

For the sake of convenience, the formula (\ref{NDoF1}) for the DoF number counting can be rewritten in the form \cite{Kaparulin:2012px}
\begin{equation}\label{NDoF2}
\displaystyle \mathcal{N}=\sum\limits_{n}n\Big(t_n-\sum\limits_{m}(-1)^m\big(l_n^m+r_n^m\big)\Big)\,,
\end{equation}
where $t_n$ is the number of equations of order $n$, and $l_n^m$ and $r_n^m$ are the numbers of gauge identities and gauge symmetries of total order $n$ and order of reducibility $m$, respectively. For example, for the Proca system (\ref{Proca}), $t_2=4$, $t_1=1$, $l_3^0=1$, and
\begin{equation}
\displaystyle \mathcal{N}=1\cdot1+2\cdot4-3\cdot1=6\,,
\end{equation}
that gives us the correct DoF number by the phase space count.

Let us now consider the linearised  Nordstr\"om
gravity defined by a single second-order equation (\ref{LNE}), $t_2=1$. The gauge parameters are represented by  traceful tensors with the symmetry of the hook type, so the number of gauge symmetry transformations (\ref{Hook1}), (\ref{Hook2})--(\ref{Hookk-}) coincides with the number of independent component of such tensors. For example, in $d=4$ case, there exist 20 first-order symmetry transformations of reducibility order 0, $r_1^0=20$, 15 second-order transformations of reducibility order 1, $r_2^1=15$, and 4 third-order transformations of reducibility order 2, $r_3^2=4$. So, one can apply (\ref{NDoF2}) and get the correct DoF number
\begin{equation}
\displaystyle \mathcal{N}=2-1\cdot20+2\cdot15-3\cdot4=0\,.
\end{equation}
For arbitrary $d$, the number of gauge symmetry transformations of order $n$ and reducibility order $n-1$ corresponds to the number of independent components of the traceful hook symmetry type tensors with $n+2$ indices, i.e.
\begin{equation}
\displaystyle r_n^{n-1}=\frac{(n+1)(d+1)!}{(n+2)!(d-n-1)!}\,,
\end{equation}
where $n=1,\ldots, d-1$. The formula (\ref{NDoF2}) reads
\begin{equation}
\displaystyle \mathcal{N}=2-\sum\limits_{n=1}^{d-1}(-1)^n\frac{n(n+1)(d+1)!}{(n+2)!(d-n-1)!}=0\,.
\end{equation}

For the dual form of the spin two theory formulated in terms of hooks (\ref{LELGH}), the formula for DoF number counting (\ref{NDoF2}) gives us the correct answer for the spin two field by phase space count, $\mathcal{N}=d^2-3d$. Let us demonstrate this computation for $d=4$.
Then, in (\ref{NDoF2}) $t_3=9$, $l^0_4=4$, $r_1^0=15$, $r^1_2=4$ (see (\ref{LELGH}), (\ref{gidLGH}), and (\ref{Hook2})--(\ref{Hookk-})), and
\begin{equation}
	\mathcal{N}=3\cdot9-4\cdot4-1\cdot15+2\cdot4=4\,,
\end{equation}
that corresponds to two polarisations.


\begin{thebibliography}{11}

\bibitem{Curtright:1980yk}
T.~Curtright, ``Generalized gauge fields'', Phys. Lett. B
\textbf{165} (1985), 304.

\bibitem{Hull:2001iu}
C.~M.~Hull, ``Duality in gravity and higher spin gauge fields'',
JHEP \textbf{09} (2001), 027
[arXiv:hep-th/0107149].

\bibitem{Casini:2001gv}
H.~Casini, R.~Montemayor and L.~F.~Urrutia,
``Dual theories for mixed symmetry fields. Spin two case: (1,1) versus (2,1) Young symmetry type fields'',
Phys. Lett. B \textbf{507} (2001), 336-344
[arXiv:hep-th/0102104].

\bibitem{Bekaert:2002uh}
X.~Bekaert, N.~Boulanger and M.~Henneaux, ``Consistent deformations of dual formulations of linearized gravity: A No go result'', Phys. Rev. D \textbf{67} (2003), 044010 [arXiv:hep-th/0210278].

\bibitem{Bekaert:2002dt}
X.~Bekaert and N.~Boulanger,
``Tensor gauge fields in arbitrary representations of GL(D,R): Duality and Poincare lemma'', Commun. Math. Phys. \textbf{245} (2004), 27-67
[arXiv:hep-th/0208058].

\bibitem{Boulanger:2020yib}
N.~Boulanger and V.~Lekeu,
``Higher spins from exotic dualisations'',
JHEP \textbf{03} (2021), 171
[arXiv:2012.11356 [hep-th]].

\bibitem{Krasnov:2021nsq}
K.~Krasnov, E.~Skvortsov, and T.~Tran,
``Actions for self-dual Higher Spin Gravities'', JHEP \textbf{08} (2021) 076 [arXiv:2105.12782 [hep-th]].

\bibitem{Lanczos:1949zz}
C.~Lanczos, ``Lagrangian Multiplier and Riemannian Spaces'', Rev. Mod. Phys. \textbf{21} (1949), 497-502.

\bibitem{Lanczos:1962zz}
C.~Lanczos, ``The Splitting of the Riemann Tensor'', Rev. Mod. Phys. \textbf{34} (1962), 379-389.

\bibitem{Takeno}
H.~Takeno, ``On the spintensor of Lanczos'', Tensor \textbf{15} (1964),  103-119.

\bibitem{Jezierski:2020ror}
J.~Jezierski, J.~Kijowski and M.~Wiatr,
``Localizing Energy in Fierz-Lanczos theory'',
Phys. Rev. D \textbf{102} (2020) no.2, 024015
[arXiv:2004.03214 [gr-qc]].

\bibitem{Gopal:2021lax}
R.~Gopal, ``Lanczos potential of Weyl field: interpretations and applications'',
Eur. Phys. J. C \textbf{81} (2021) no.2, 194 [arXiv:2103.00311 [gr-qc]].

\bibitem{Edgar:1997af}
S.~B.~Edgar and A.~Hoglund,
``The Lanczos potential for Weyl candidate tensors exists only in four-dimensions'',
Gen. Rel. Grav. \textbf{32} (2000), 2307-2318
[arXiv:gr-qc/9711022].

\bibitem{Edgar:2004iq}
S.~B.~Edgar and J.~M.~M.~Senovilla,
``A Local potential for the Weyl tensor in all dimensions'',
Class. Quant. Grav. \textbf{21} (2004), L133
[arXiv:gr-qc/0408071].

\bibitem{Bergshoeff:2009tb}
E.~A.~Berghoeff, O.~Hohm and P.~K.~Townsend, ``On Higher Derivatives in 3D Gravity and Higher Spin Gauge Theories'', Annals Phys. \textbf{325} (2010), 1118-1134 [arXiv:0911.3061 [hep-th]].

\bibitem{Abakumova:2021evc}
V.~A.~Abakumova, S.~L.~Lyakhovich,
``Reducible Stueckelberg symmetry and dualities'',
Phys. Lett. B \textbf{820} (2021), 136552
[arXiv:2106.09355 [hep-th]].

\bibitem{Koiso} N.~Koiso, ``A decomposition of the space $\mathcal{M}$  of Riemannian metrics on a manifold'', Osaka J. Math. \textbf{16}(2) (1979), 423-429.

\bibitem{Besse:1987pua} Arthur L. Besse,  ``Einstein Manifolds'', Berlin: Springer, 1977.

\bibitem{Lyakhovich:2008hu}
S.~L.~Lyakhovich and A.~A.~Sharapov,
``Normal forms and gauge symmetries of local dynamics'',
J. Math. Phys. \textbf{50} (2009), 083510
[arXiv:0812.4914 [math-ph]].

\bibitem{Lyakhovich:2013qfa}
S.~L.~Lyakhovich and A.~A.~Sharapov,
``Gauge symmetries in 2D field theory'',
J. Geom. Phys. \textbf{97} (2015), 227-242
[arXiv:1312.2671 [math-ph]].

\bibitem{Francia:2013sca}
D.~Francia, S.~L.~Lyakhovich and A.~A.~Sharapov,
``On the gauge symmetries of Maxwell-like higher-spin Lagrangians'',
Nucl. Phys. B \textbf{881} (2014), 248-268
[arXiv:1310.8589 [hep-th]].

\bibitem{E} D. Eisenbud,  ``The geometry of syzygies. A second course in commutative algebra and algebraic geometry''.
Graduate Texts in Mathematics 229, New York: Springer-Verlag, 2005.

\bibitem{M2}
D.~R.~Grayson and Michael~E.~Stillman,
``Macaulay2, a software system for research in algebraic geometry'',
Available at \url{http://www.math.uiuc.edu/Macaulay2/}.

\bibitem{Kaparulin:2012px}
D.~S.~Kaparulin, S.~L.~Lyakhovich and A.~A.~Sharapov,
``Consistent interactions and involution'',
JHEP \textbf{01} (2013), 097
[arXiv:1210.6821 [hep-th]].

\bibitem{AL2021}
V.~A.~Abakumova, S.~L.~Lyakhovich, ``Hamiltonian constrained formalism for the general field theories with unfree gauge symmetry'', Memoirs of the Faculty of Physics (2021) no. 1, 2111503 [in Russian].

\bibitem{Lyakhovich:2014soa}
S.~L.~Lyakhovich, A.~A.~Sharapov,
``Multiple choice of gauge generators and consistency of interactions'',
Mod. Phys. Lett. A \textbf{29} (2014) no.31, 1450167
[arXiv:1402.0634 [gr-qc]].

\bibitem{Lyakhovich:2021lzy}
S.~L.~Lyakhovich,
``General method for including Stueckelberg fields,''
Eur. Phys. J. C \textbf{81} (2021) no.5, 472
[arXiv:2102.10579 [hep-th]].

\bibitem{Kaparulin:2014vpa}
D.~S.~Kaparulin, S.~L.~Lyakhovich and A.~A.~Sharapov,
``Classical and quantum stability of higher-derivative dynamics'',
Eur. Phys. J. C \textbf{74} (2014) no.10, 3072
[arXiv:1407.8481 [hep-th]].

\bibitem{Abakumova:2018eck}
V.~A.~Abakumova, D.~S.~Kaparulin and S.~L.~Lyakhovich,
``Stable interactions in higher derivative field theories of derived type'',
Phys. Rev. D \textbf{99} (2019) no.4, 045020
[arXiv:1811.10019 [hep-th]].

\bibitem{Abakumova:2017uto}
V.~A.~Abakumova, D.~S.~Kaparulin and S.~L.~Lyakhovich,
``Multi-Hamiltonian formulations and stability of higher-derivative extensions of $3d$ Chern-Simons'',
Eur. Phys. J. C \textbf{78} (2018) no.2, 115
[arXiv:1711.07897 [hep-th]].

\end{thebibliography}
\end{document}